# Ideal quantum clocks and operator time

Walter Gessner
Unterer Katzenbergweg 7
D-97084 Wuerzburg

## Abstract

In the framework of any quantum theory in the Schrödinger picture a general operator time concept is given. For this purpose special closed systems are emphasized as "ideal quantum clocks" (IQC's). Their definition follows heuristically from a common property of ideal clocks and from general postulates of traditional quantum theory. It is assumed that any IQC C is described by a special solution $\varphi_C(t)$ of Schrödingers equation in a suitable state space S and that an infinite set of equidistant time points $\tau^n$ exists so that the elements $\varphi_C(\tau^n)$ are pairwise orthogonal and span a certain time invariant subspace $s_C$ of S. On $s_C$, which contains the whole curve $\varphi_C(t)$, a symmetric time operator $T_C$ is defined. $T_C$ and the Hamiltonian $H_C$ of the IQC satisfies $[T_C, H_C] = i$ so that the time-energy uncertainty relation holds. Further it is discussed how the IQC interacts with another physical system D and to what extent the IQC measures the time differences of any prescribed initial and final states of D in spite of the disturbance of C by D.

## 1. From ideal to quantum clocks.

In this paper a further attempt to construct a time operator in the traditional quantum theoretical framework (QM, QED, Schrödinger picture) is given, but on the basis of a general concept of ideal quantum clocks.

In the quantum theoretical framework any closed physical system corresponds an element φ of a certain pre-Hilbert or Hilbert space S (scalar product $<…|…>$, norm $\|φ\|$), and a Hamiltonian H which determines the time evolution φ(t) of φ by $U(t)φ = \exp(-itH)φ$, $t \in \mathbb{R}$. The elements φ(t), $t \in \mathbb{R}$, build a set of elements of S, called "Schrödinger curve".

To find now the concept of an IQC, common properties of ideal clocks are translated via general postulates of quantum theory.

Any clock has a certain time resolution $τ > 0$ so that the reading of the clock provides one of the values of the set $\{0, \pm τ, \pm 2τ, …\}$. This set consists of equidistant time points $τ^n := nτ$, $n \in \mathbb{Z} = \{0, \pm 1, \pm 2, …\}$, collected in the sequence $<τ^n>$. The best clocks of today satisfy $τ \approx 10^{-15}$ s. A clock with $<τ^n>$ defines a set of "states" $Z^n$, the "clicks", standing in a one-to-one correspondence to the $τ^n$. For example, $Z^n$ may be a number sequence in the dial of a digital clock, a special mode of the resonator of a cesium atomic watch, the number of atoms in the ground state in an ensemble of (at t = 0) excited atoms, and arises when and only when the running time parameter t reaches $τ^n$. Vice versa the clock is read off by taking knowledge of $Z^n$ and concluding from $Z^n$ to the time point $τ^n$ with full certainty.

Because the time measurement is assumed to be an intrinsic element of the quantum theory in question, any IQC to be introduced is a closed physical system with a Schrödinger curve $φ_C(t) = \exp(-itH_C)φ_C$, in a certain pre-Hilbert or Hilbert state space S. The Hamiltonian $H_C$ determines the time evolution of the IQC. The role of $Z^n$ is to be taken by $φ_C(τ^n)$ so that these $φ_C(τ^n)$ are now the clicks of the IQC. Whereas any $Z^n$ delivers the sharp time point $τ^n$, the corresponding $φ_C(τ^n)$ will deliver only an expectation value with a nonvanishing standard deviation of a time operator to be defined. Because $Z^m$ and $Z^n$ exclude one another for m ≠ n, the probabilities $|\langle\varphi_C(\tau^m)|\varphi_C(\tau^n)\rangle|^2$ have to vanish for all pairwise different m, n. As a consequence, a set of pairwise orthonormal elements $\{\varphi_C(\tau^n)|n \in \mathbb{Z}\}$ does exist so that the first choice of a suitable state space of C is the Hilbert space

$$S_C := \text{span}\{\varphi_C(\tau^n)|n \in \mathbb{Z}\}. \tag{01}$$

But for reasons of convergence (section 4) the subspace $s_C$ of $S_C$ of all $\varphi := \sum_{n=-\infty}^{+\infty} d^n \varphi_C(\tau^n)$ with complex valued $d^n$ satisfying $\sum_{n=-\infty}^{+\infty} |n||d^n| < \infty$, resp.

$$s_C := \{\varphi \in S_C | \sum_{n=-\infty}^{+\infty} |n||\langle\varphi_C(\tau^n)|\varphi\rangle| < \infty \}, \tag{02}$$

is to be introduced. The pre-Hilbert space $s_C$ is dense in $S_C$. To guarantee $H_C\varphi_C(t) \in s_C$ for all t (Lemma 1), it is assumed with respect to Schrödinger's equation that $d/dt\, \varphi_C(0) \in s_C$.

The $\varphi_C(\tau^n)$ are elements of one and the same Schrödinger curve so that $U(t)\varphi_C(\tau^n) = \varphi_C(t + \tau^n)$ for all n. To any fixed t, $\{\varphi_C(t + \tau^n)|n \in \mathbb{Z}\}$ is an orthonormal basis of $U(t)s_C \subseteq S$. The common time parameter t in the $\varphi_C(t + \tau^n)$ provides only a translation of the $\tau^n$ along the t-axis and thus a special "setting" of C so that the bases $\{\varphi_C(t + \tau^n)|n \in \mathbb{Z}\}$ and $\{\varphi_C(\tau^n)|n \in \mathbb{Z}\}$ define the same C. Because the only setting of an IQC should not determine its state space it is assumed, that the $\varphi_C(t + \tau^n)$ also span $s_C$ resp. $S_C$ defined initially by the $\varphi_C(\tau^n)$. This reads $U(t)s_C = s_C$ for all t (so that $U(t)S_C = S_C$). Then $s_C$ and $S_C$ are time invariant and $s_C$ contains the whole curve $\varphi_C(t)$. Because now $U(t)s_C = s_C$, the $\varphi_C(\tau^n)$ in (02) can be replaced by $\varphi_C(t + \tau^n)$ so that $s_C$ is no longer defined by a special basis. Summarizing, the freedom of setting the IQC leads to the condition $U(t)s_C = s_C$ for all t.

As a consequence, $\varphi_C(t + \tau^n)$ can be written as $\varphi_C(t + \tau^n) = \sum_{m=-\infty}^{+\infty} c^{mn}(t)\varphi_C(\tau^m)$ for all n with complex valued $c^{mn}(t)$. These C-characteristic functions $c^{mn}(t)$ provide (section 3) tools for the following discussions.

In [1] a concept of quantum clocks with a finite set $\langle\tau^n\rangle$ is introduced, and some examples of such clocks are discussed although with different aims than constructing a time operator. An explicit quantum clock is the oscillator clock of [2] in which a bounded and selfadjoint time operator (canonically conjugated to the oscillator Hamiltonian) is used. In [4] the quantum clock concept of [1] is taken and an operator similar to the $P_C$ of section 4, equ. (21), is used as a time operator. But the aims of [4] are the couplings of a clock to another system, the construction of the corresponding Hamiltonians as well as the discussion of the energy exchanges between the clock and the other system. The different meanings of the multifaceted term "time" are treated in a theory of measurement [8, 10-12]. With respect to the "arrival time" self-adjoint operators are constructed resp. probability distributions are defined [13, 14]. The detailed analysis [15] of the objection of Pauli [5 - 7] against any time operator shows that by careful regard of the domains of the operators involved there may exist a class of selfadjoint time operators canonically conjugate to a given semibounded Hamiltonian. In [16] a time operator, appropriate to define the time of passage or arrival time at a specific point, is introduced. It is shown that the uncertainty relation is satisfied and that Pauli's objection can be resolved or circumvented. Especially to circumvent this objection, in [17, 18] instead of selfadjoint time operators positive operator valued measures are used.

The IQC-theory of this paper may revive time operators in the traditional quantum theory. It seems that up to date no similar concepts are published.

## 2. Ideal quantum clocks.

So to call, "clicks and setting" of a clock lead now to the concept of an IQC:

**Definition 1:**
A closed system C is an **ideal quantum clock** (IQC), if a Hilbert resp. pre-Hilbert space S, a Schrödinger curve $\varphi_C(t) \in S$ with $d/dt\, \varphi_C(0) \in s_C$ and a time interval $\tau > 0$ exists so that ($\tau^n := n\tau,\ n \in \mathbb{Z}$)

i) $\langle \varphi_C(\tau^m) | \varphi_C(\tau^n) \rangle = \delta^{mn}$ for all $m, n \in \mathbb{Z}$. (03)

ii) $s_C$ is time invariant: $U(t) s_C = s_C$ for all $t$. (04)

The time interval $\tau$ cannot be arbitrarily small. The transition $\tau \to 0$ namely causes a lot of troubles, as especially in [1] is pointed out. At last, to avoid the vanishing of $\tau$ in the "noise" of the time caused by the graviton field, $\tau$ very much has to exceed the Planck time.

Examples:
- The number of atoms in the ground state in an assemble of (at $t = 0$) excited atoms; the clicks are the transitions of excited atoms into the ground state. Electric circuits: the clicks are the transitions of a fixed charge q from a certain component A to a component B of the circuit at equidistant time points $\tau^n$.
- The examples discussed in [1, 2, 3].

Obviously, all these examples are not IQC's in the rigorous sense (03, 04), but approximate the IQC-concept to some extent. It is an open question whether the exact Schrödinger term of a special IQC can be found or not. But definition 1 follows from general properties of ideal clocks and from general principles of quantum theory and gives insofar the quintessence of an ideal quantum clock.

## 3. The C-characteristic functions.

Furtheron a Schrödinger curve $\varphi_C(t) \in s_C$ for all t is assumed. Instead on the whole t-line, $\varphi_C(t)$ is now discussed on the t-intervals $[\tau^k - \tau/2; \tau^k + \tau/2]$ for all $k \in \mathbb{Z}$. The restriction of $\varphi_C(t)$ to any of these intervals reads $\varphi_C(\tau^k + u)$ with $u \in [-\tau/2; +\tau/2]$ and can as an element from $s_C$ be written as

$$\varphi_C(\tau^k + u) = \sum_{n=-\infty}^{+\infty} c^{nk}(u) \varphi_C(\tau^n) \tag{05}$$

with C-characteristic complex valued functions $c^{nk}(u)$.

In the following some properties of the $c^{nk}(u) = \langle \varphi_C(\tau^n) | \varphi_C(\tau^k + u) \rangle$ are needed. All indices are from $\mathbb{Z}$ and $u \in [-\tau/2; +\tau/2]$. The conjugate of the complex valued c is denoted by $c^*$:

a) $c^{n+r\ k+r}(u) = c^{nk}(u)$. (06)

Proof: $c^{n+r\ k+r}(u) = \langle \varphi_C(\tau^{n+r}) | \varphi_C(\tau^{k+r} + u) \rangle = \langle \varphi_C(\tau^n) | U^+(\tau^r) U(\tau^r) | \varphi_C(\tau^k + u) \rangle = \langle \varphi_C(\tau^n) | \varphi_C(\tau^k + u) \rangle = c^{nk}(u)$.

b) $c^{mn}(-u) = c^{nm}(u)^*$. (07)

Proof: $c^{mn}(-u) = \langle \varphi_C(\tau^m) | U(-u) \varphi_C(\tau^n) \rangle = \langle \varphi_C(\tau^m) U^+(u) | \varphi_C(\tau^n) \rangle = \langle \varphi_C(\tau^m + u) | \varphi_C(\tau^n) \rangle = \langle \varphi_C(\tau^n) | \varphi_C(\tau^m + u) \rangle^* = c^{nm}(u)^*$.

c) $\sum_{n=-\infty}^{+\infty} c^{nk}(u)^* c^{nm}(u) = \delta^{km} = \sum_{n=-\infty}^{+\infty} c^{kn}(u)^* c^{mn}(u)$, especially (08)

$$\sum_{n=-\infty}^{+\infty} |c^{nk}(u)|^2 = 1 = \sum_{n=-\infty}^{+\infty} |c^{kn}(u)|^2.$$

This follows from $\langle \varphi_C(\tau^k + u) | \varphi_C(\tau^m + u) \rangle = \delta^{km}$ and (08).

Because of $\varphi_C(\tau^k + u) \in s_C$ and the definition of $s_C$, the $c^{nk}(u)$ satisfy the convergence condition (02)

$$\sum_{n=-\infty}^{+\infty} |n| |c^{nk}(u)| < \infty \text{ for any } u \in [-\tau/2; +\tau/2]. \tag{09}$$

The derivations of the $c^{nk}(u)$ satisfy analogous convergence conditions:

d/dt $\varphi_C(0) \in s_C$ reads $\sum_{n=-\infty}^{+\infty} \dot{c}^{n0}(0)\varphi_C(\tau^n) \in s_C$ where $\dot{c}^{n0}(0) := d/du\, c^{n0}(0)$ so that

$$\sum_{n=-\infty}^{+\infty} |n||\dot{c}^{n0}(0)| < \infty. \tag{10}$$

From this follows for all $k \in \mathbb{Z}$

$$\sum_{n=-\infty}^{+\infty} |n||\dot{c}^{nk}(0)| < \infty \text{ and } \sum_{n=-\infty}^{+\infty} |n||\dot{c}^{kn}(0)| < \infty, \tag{11}$$

Proof:

$$\sum_{n=-\infty}^{+\infty} |n||\dot{c}^{nk}(0)| = \sum_{n=-\infty}^{+\infty} |(n-k)+k||\dot{c}^{n-k\,0}(0)| \leq \sum_{n=-\infty}^{+\infty} |n-k||\dot{c}^{n-k\,0}(0)| +$$

$$+ |k| \sum_{n=-\infty}^{+\infty} |\dot{c}^{n-k\,0}(0)| < \infty \text{ to any fixed k according to (10).}$$

$$\sum_{n=-\infty}^{+\infty} |n||\dot{c}^{kn}(0)| = \sum_{n=-\infty}^{+\infty} |n||\dot{c}^{nk}(0)| < \infty \text{ by (10). A similar conclusion shows}$$

that (11) is a consequence of $\sum_{n=-\infty}^{+\infty} |n||c^{n0}(u)| < \infty$.

Thus, the definition of an IQC contains implicitly the convergence properties (11).

**Lemma 1:**

a) $H_C s_C \subseteq s_C$; $H_C S_C \subseteq S_C$. $\hfill(12)$

b) $\langle \varphi_C(t) | H_C | \varphi_C(t) \rangle = i\dot{c}^{00}(0)$ for all $t \in \mathbb{R}$.

c) $|\langle \varphi | H_C | \varphi \rangle| \leq W < \infty$ for all normed $\varphi \in S_C$, $\hfill(13)$

with $W := \sum_{r=-\infty}^{+\infty} |\dot{c}^{r\,0}(0)|$.

**Proof:**

a) To prove $H_C \varphi \in s_C$ if $\varphi = \sum_{n=-\infty}^{+\infty} d^n \varphi_C(\tau^n) \in s_C$ one gets first

$H_C \varphi = i \sum_{m=-\infty}^{+\infty} \sum_{n=-\infty}^{+\infty} d^n \dot{c}^{mn}(0) \varphi_C(\tau^m)$. From (11) follows immediately

$$\sum_{m=-\infty}^{+\infty} \sum_{n=-\infty}^{+\infty} |m||d^n||\dot{c}^{mn}(0)| < \infty \text{ so that } H_C \varphi \in s_C. \tag{14}$$

To prove $H_C S_C \subseteq S_C$ one gets first $U(u)\varphi = \sum_{m=-\infty}^{+\infty} \sum_{n=-\infty}^{+\infty} d^n c^{mn}(u)\varphi_C(\tau^m)$ where

$\varphi = \sum_{n=-\infty}^{+\infty} d^n \varphi_C(\tau^n) \in S_C$, so that with $H_C\varphi = i[d/du\, U(u)\varphi](0)$

$$H_C\varphi = i \sum_{m=-\infty}^{+\infty} \sum_{n=-\infty}^{+\infty} d^n \dot{c}^{mn}(0)\varphi_C(\tau^m), \tag{15}$$

furtheron

$$\|H_C\varphi\|^2 = \sum_{m=-\infty}^{+\infty} \sum_{n=-\infty}^{+\infty} \sum_{r=-\infty}^{+\infty} (d^r)^* d^m \dot{c}^{nr}(0)^* \dot{c}^{nm}(0) = \tag{16}$$

$$\sum_{m=-\infty}^{+\infty} \sum_{n=-\infty}^{+\infty} \sum_{r=-\infty}^{+\infty} (d^r)^* d^m \dot{c}^{0\, r-n}(0)^* \dot{c}^{0\, m-n}(0) =$$

$$\sum_{a=-\infty}^{+\infty} \sum_{b=-\infty}^{+\infty} \sum_{n=-\infty}^{+\infty} (d^{a+n})^* d^{b+n} \dot{c}^{0\, a}(0)^* \dot{c}^{0\, b}(0) \leq$$

$$\sum_{a=-\infty}^{+\infty} \sum_{b=-\infty}^{+\infty} |\dot{c}^{0\, a}(0)||\dot{c}^{0\, b}(0)|| \sum_{n=-\infty}^{+\infty} (d^{(a-b)+n})^* d^n |, \text{ where } a := r-n,\ b := m-n.$$

$$|\sum_{n=-\infty}^{+\infty} (d^{(a-b)+n})^* d^n|^2 = \tag{17}$$

$$|\langle\varphi|U(\tau^{a-b})\varphi\rangle|^2 \leq \langle\varphi|\varphi\rangle\langle U(\tau^{a-b})\varphi|U(\tau^{a-b})\varphi\rangle = 1$$

so that $\|H_C\varphi\|^2 \leq \sum_{a=-\infty}^{+\infty} \sum_{b=-\infty}^{+\infty} |\dot{c}^{0\, a}(0)||\dot{c}^{0\, b}(0)| < \infty$ because of (11).

Thus $H_C\varphi \in S_C$.

b) $H_C\varphi_C(\tau^k+u) = i \sum_{n=-\infty}^{+\infty} \dot{c}^{nk}(u)\varphi_C(\tau^n)$ so that

$$\langle\varphi_C(\tau^k+u)|H_C|\varphi_C(\tau^k+u)\rangle = i \sum_{n=-\infty}^{+\infty} c^{nk}(u)^* \dot{c}^{nk}(u) = \langle\varphi_C(\tau^0)|H_C|\varphi_C(\tau^0)\rangle = i\dot{c}^{00}(0)$$

because the expectation value of $H_C$ is time independent, and $c^{nk}(0) = \delta^{nk}$.

c) $H_C\varphi = i \sum_{m=-\infty}^{+\infty} \sum_{n=-\infty}^{+\infty} d^n \dot{c}^{mn}(0)\varphi_C(\tau^m) \in S_C$, so that $\tag{18}$

$$\langle\varphi|H_C|\varphi\rangle = \sum_{m=-\infty}^{+\infty} \sum_{n=-\infty}^{+\infty} i(d^m)^* d^n \dot{c}^{mn}(0) = \sum_{m=-\infty}^{+\infty} \sum_{n=-\infty}^{+\infty} i(d^m)^* d^n \dot{c}^{m-n\, 0}(0) =$$

$$\sum_{m=-\infty}^{+\infty} \sum_{n=-\infty}^{+\infty} i(d^{(m-n)+n})^* d^n \dot{c}^{m-n\,0}(0) = \sum_{n=-\infty}^{+\infty} \sum_{r=-\infty}^{+\infty} i(d^{r+n})^* d^n \dot{c}^{r\,0}(0).$$

Therefore,

$$|\langle\varphi|H_C|\varphi\rangle| \leq \sum_{r=-\infty}^{+\infty} |\dot{c}^{r\,0}(0)| |\sum_{n=-\infty}^{+\infty} (d^{r+n})^* d^n|. \qquad (19)$$

In analogy to (17) one has

$$|\sum_{n=-\infty}^{+\infty} (d^{r+n})^* d^n|^2 = |\langle\varphi|U(\tau^r)\varphi\rangle|^2 \leq \langle\varphi|\varphi\rangle\langle\varphi|U(\tau^r)^+ U(\tau^r)|\varphi\rangle = 1, \qquad (20)$$

for any $r \in \mathbb{Z}$ so that $|\langle\varphi|H_C|\varphi\rangle| \leq \sum_{r=-\infty}^{+\infty} |\dot{c}^{r\,0}(0)| < \infty$. ∎

## 4. The time operator $T_C$ of C.

It is now discussed how to read C. First, the mapping $P_C: s_C \to S_C$ orders to any $\varphi := \sum_{n=-\infty}^{+\infty} d^n \varphi_C(\tau^n) \in s_C$ all those $\tau^n$ with corresponding probabilities whose $\varphi_C(\tau^n)$ arise in the expansion of $\varphi$:

$$P_C \varphi := \sum_{n=-\infty}^{+\infty} d^n \tau^n \varphi_C(\tau^n). \tag{21}$$

$\|P_C \varphi\| \leq \tau \sum_{n=-\infty}^{+\infty} |n| |d^n| < \infty$, so that $P_C \varphi \in S_C$.

As a time operator, $P_C$ should at least satisfy $\langle \varphi_C(t) | P_C | \varphi_C(t) \rangle = t$, so that $d/dt \langle \varphi_C(t) | P_C | \varphi_C(t) \rangle = -i \langle \varphi_C(t) | [P_C, H_C] | \varphi_C(t) \rangle = 1$ for all t. Instead, one has $\langle \varphi_C(\tau^n) | [P_C, H_C] | \varphi_C(\tau^n) \rangle = 0$, because all $\varphi_C(\tau^n)$ are eigenstates of $P_C$. Therefore, $P_C$ is no time operator. But in the next section it is shown that a suitable averaging of $P_C$ leads to an operator with the desired properties. This requires the lemma:

**Lemma 2:**

Let $\varphi \in S_C$, $\|\varphi\| = 1$ and $\varphi(t) = U(t)\varphi \in S_C$ (by (12)). Then

$$\langle \varphi(+\tau/2) | P_C | \varphi(+\tau/2) \rangle = \tau + \langle \varphi(-\tau/2) | P_C | \varphi(-\tau/2) \rangle. \tag{22}$$

**Proof:**

As an element of $S_C$, $U(-\tau/2)\varphi$ can be written as $\sum_{n=-\infty}^{+\infty} d^n \varphi_C(\tau^n)$ with complex valued $d^n$. One gets then $U(\tau) P_C [U(-\tau/2)\varphi] = \sum_{n=-\infty}^{+\infty} d^n \tau^n U(\tau) \varphi_C(\tau^n). \tag{23}$

On the other hand

$$P_C U(\tau)[U(-\tau/2)\varphi] = \sum_{n=-\infty}^{+\infty} d^n P_C U(\tau) \varphi_C(\tau^n) = \sum_{n=-\infty}^{+\infty} d^n P_C \varphi_C(\tau^{n+1}) =$$

$$\sum_{n=-\infty}^{+\infty} d^n \tau^{n+1} \varphi_C(\tau^{n+1}) = \sum_{n=-\infty}^{+\infty} d^n (\tau + \tau^n) \varphi_C(\tau^{n+1}) = \sum_{n=-\infty}^{+\infty} d^n (\tau + \tau^n) U(\tau) \varphi_C(\tau^n). \tag{24}$$

Subtraktion (24) – (23) yields

$P_C U(\tau) U(-\tau/2)\varphi - U(\tau) P_C U(-\tau/2)\varphi = \tau U(\tau) \sum_{n=-\infty}^{+\infty} d^n \varphi_C(\tau^n) = \tau U(\tau)[U(-\tau/2)\varphi]$

so that

$$P_C U(+\tau/2)\varphi = \tau U(+\tau/2)\varphi + U(\tau)P_C U(-\tau/2)\varphi. \tag{25}$$

The scalar product of (25) and $U(+\tau/2)\varphi$ leads to

$$\langle \varphi U^+(+\tau/2)|P_C U(+\tau/2)\varphi\rangle = \tau + \langle \varphi U^+(+\tau/2)U^+(-\tau)|P_C U(-\tau/2)\varphi\rangle$$

so that $\langle \varphi(+\tau/2)|P_C|\varphi(+\tau/2)\rangle = \tau + \langle \varphi(-\tau/2)|P_C|\varphi(-\tau/2)\rangle$. ∎

For the following definition it is to be emphasized that the Hamiltonian $H_C$, the time evolution $U(t)$ and $P_C$ satisfy

$H_C: S_C \to S_C$, $H_C: s_C \to s_C$, $U(t): S_C \to S_C$, $U(t): s_C \to s_C$, and $P_C: s_C \to S_C$, so that especially the term $U^+(u)P_C U(u): s_C \to S_C$ in (26) is well defined. The time operator is now given by an averaging of $P_C$ along the section of the Schrödinger curve $U(u)\varphi \in s_C$, $\varphi \in s_C$, $u \in [-\tau/2; +\tau/2]$.

**Definition 2:**

The time operator $T_C: s_C \to S_C$ of C is introduced by

$$T_C := \tau^{-1} \int_{-\tau/2}^{+\tau/2} du\, U^+(u)P_C U(u). \tag{26}$$

In Lemma 3 the existence of $T_C$ as a mapping $T_C: s_C \to S_C$ is proven and more familiar terms of $T_C$ are given (27, 37).

**Lemma 3:**

The domain of $T_C$ is $s_C$. To any $\varphi = \sum_{k=-\infty}^{+\infty} d^k \varphi_C(\tau^k) \in s_C$ one gets

$$T_C \varphi = \sum_{k=-\infty}^{+\infty} \sum_{m=-\infty}^{+\infty} d^k\, C^{km} \varphi_C(\tau^m), \text{ especially} \tag{27}$$

$$T_C \varphi_C(\tau^k) = \sum_{m=-\infty}^{+\infty} C^{km} \varphi_C(\tau^m), \text{ with} \tag{28}$$

$$C^{km} := \tau^{-1} \sum_{n=-\infty}^{+\infty} \int_{-\tau/2}^{+\tau/2} du\, c^{kn}(u)^* \tau^n\, c^{mn}(u), \tag{29}$$

where the $c^{mn}(u)$ are the C-characteristic functions introduced by (05). The $C^{km}$ satisfy $(C^{km})^* = C^{mk}$.

**Proof:**

At first, $T_C\varphi_C(\tau^k) = \tau^{-1} \int_{-\tau/2}^{+\tau/2} du\, U^+(u) P_C \varphi_C(\tau^k + u)$. By (05) one gets

$$T_C\varphi_C(\tau^k) = \tau^{-1} \sum_{n=-\infty}^{+\infty} \int_{-\tau/2}^{+\tau/2} du\, c^{nk}(u) U^+(u) P_C \varphi_C(\tau^n) = \quad (30)$$

$$\tau^{-1} \sum_{n=-\infty}^{+\infty} \int_{-\tau/2}^{+\tau/2} du\, c^{nk}(u)\, \tau^n\, U^+(u) \varphi_C(\tau^n) = \tau^{-1} \sum_{n=-\infty}^{+\infty} \int_{-\tau/2}^{+\tau/2} du\, c^{nk}(u)\, \tau^n\, \varphi_C(\tau^n - u). \quad (31)$$

$\varphi_C(\tau^n - u)$ with $u \in [-\tau/2; +\tau/2]$ can be written by (05) as

$$\varphi_C(\tau^n - u) = \sum_{m=-\infty}^{+\infty} c^{nm}(u)^* \varphi_C(\tau^m) \text{ so that}$$

$$T_C\varphi_C(\tau^k) = \tau^{-1} \sum_{m=-\infty}^{+\infty} \sum_{n=-\infty}^{+\infty} \int_{-\tau/2}^{+\tau/2} du\, c^{nk}(u)\, \tau^n\, c^{nm}(u)^* \varphi_C(\tau^m) =$$

$$\tau^{-1} \sum_{m=-\infty}^{+\infty} \sum_{n=-\infty}^{+\infty} \int_{-\tau/2}^{+\tau/2} du\, c^{nk}(-u)\, \tau^n\, c^{nm}(-u)^* \varphi_C(\tau^m) =$$

$$\tau^{-1} \sum_{m=-\infty}^{+\infty} \sum_{n=-\infty}^{+\infty} \int_{-\tau/2}^{+\tau/2} du\, c^{kn}(u)^*\, \tau^n\, c^{mn}(u) \varphi_C(\tau^m). \quad (32)$$

Therefore,

$$T_C\varphi_C(\tau^k) = \sum_{m=-\infty}^{+\infty} C^{km}\, \varphi_C(\tau^m), \text{ where}$$

$$C^{km} := \tau^{-1} \sum_{n=-\infty}^{+\infty} \int_{-\tau/2}^{+\tau/2} du\, c^{kn}(u)^*\, \tau^n\, c^{mn}(u).$$

It remains to show that $T_C\varphi = \sum_{k=-\infty}^{+\infty} \sum_{m=-\infty}^{+\infty} d^k\, C^{km} \varphi_C(\tau^m) \in S_C$ if $\varphi \in s_C$.

At first one has $\|T_C\varphi\| \leq \sum_{k=-\infty}^{+\infty} \sum_{m=-\infty}^{+\infty} |d^k||C^{km}|. \quad (33)$

$\sum_{m=-\infty}^{+\infty} |C^{km}| \leq \int_{-\tau/2}^{+\tau/2} du \sum_{m=-\infty}^{+\infty} \sum_{n=-\infty}^{+\infty} |c^{kn}(u)||n||c^{mn}(u)|$. To any fixed n the term

$a(u) := \sum_{m=-\infty}^{+\infty} |c^{mn}(u)| = \sum_{m=-\infty}^{+\infty} |c^{m-n\, 0}(u)|$ is finite and does not depend on n. It remains

$$\sum_{m=-\infty}^{+\infty} |C^{km}| \leq \int_{-\tau/2}^{+\tau/2} du\, a(u) \sum_{n=-\infty}^{+\infty} |c^{kn}(u)||n| = \int_{-\tau/2}^{+\tau/2} du\, a(u) \sum_{n=-\infty}^{+\infty} |c^{0\,n-k}(u)||(n-k)+k|.$$

With $b(u) := \sum_{n=-\infty}^{+\infty} |c^{0\,n-k}(u)||n-k|$ and $c(u) := \sum_{n=-\infty}^{+\infty} |c^{0\,n-k}(u)|$ which are finite because of (11) and independent of k one gets

$$\sum_{m=-\infty}^{+\infty} |C^{km}| \leq \int_{-\tau/2}^{+\tau/2} du\, a(u)[b(u) + |k|c(u)] \quad \text{so that} \quad \sum_{m=-\infty}^{+\infty} |C^{km}| \leq a + b|k| \quad (34)$$

with certain $a, b \in \mathbb{R}$. Summarizing one gets

$$\|T_C \varphi\| \leq \sum_{k=-\infty}^{+\infty} |d^k|(a + b|k|) \quad \text{which is finite because of (02).} \quad (35)$$

Therefore, the domain of $T_C$ is $s_C$ so that $T_C: s_C \to S_C$. ∎

As a consequence, the expectation value $\langle \varphi | T_C \varphi \rangle$ of

$$\varphi = \sum_{n=-\infty}^{+\infty} d^n \varphi_C(\tau^n) \in s_C \quad \text{reads} \quad \langle \varphi | T_C \varphi \rangle = \sum_{k=-\infty}^{+\infty} \sum_{m=-\infty}^{+\infty} d^k C^{km} (d^m)^*. \quad (36)$$

The expectation value $\langle \varphi | T_C \varphi \rangle$ allows a more familiar term:

Because $U(u)\varphi$ can be written as $\sum_{n=-\infty}^{+\infty} d^n(u)\varphi_C(\tau^n)$ with complex valued $d^n(u)$

one gets $\langle \varphi | T_C \varphi \rangle = \tau^{-1} \sum_{n=-\infty}^{+\infty} \tau^n \int_{-\tau/2}^{+\tau/2} du\, |d^n(u)|^2.$ (37)

The time operator $T_C: s_C \to S_C$ has now the expected properties:

**Theorem:**

a) $T_C$ is symmetric.

b) $T_C$ and $H_C$ are canonically conjugated on the pre-Hilbert space $s_C$:

   $[T_C, H_C] = i$, in the integral form ($t \in \mathbb{R}$, $U(t) = \exp(-itH_C)$)  (38)

   $[T_C, U(t)] = tU(t)$.

c) $\langle \varphi(t) | T_C | \varphi(t) \rangle = t + \langle \varphi(0) | T_C | \varphi(0) \rangle$,  (39)

   if $\varphi(t) = U(t)\varphi$, $\varphi \in s_C$, $\|\varphi\| = 1$, especially

   $\langle \varphi_C(t) | T_C | \varphi_C(t) \rangle = t$.

d) The standard deviations $\sigma(T_C)$ and $\sigma(H_C)$ are time invariant

and satisfy the time-energy uncertainty relation

$$\sigma(T_C)\sigma(H_C) \geq \tfrac{1}{2}, \tag{40}$$

where $\sigma(T_C)$, and accordingly $\sigma(H_C)$, is given by ($\varphi \in s_C$)

$$\sigma(T_C)^2 = \text{Var}(T_C) = <T_C\varphi|T_C\varphi> - <\varphi|T_C|\varphi>^2. \tag{41}$$

**Proof:**

a) Because of (27) one has for $\varphi = \sum\limits_{n=-\infty}^{+\infty} d^n \varphi_C(\tau^n)$ and $\psi = \sum\limits_{n=-\infty}^{+\infty} e^n \varphi_C(\tau^n)$,

$\varphi, \psi \in s_C$, the terms $T_C\varphi = \sum\limits_{m=-\infty}^{+\infty} \sum\limits_{n=-\infty}^{+\infty} d^m C^{mn} \varphi_C(\tau^n) \in S_C$, analogously

for $T_C\psi$, so that as a consequence of $(C^{mn})^* = C^{nm}$

$$<T_C\varphi|\psi> = \sum\limits_{m=-\infty}^{+\infty} \sum\limits_{n=-\infty}^{+\infty} (d^m)^* (C^{mn})^* e^n = \sum\limits_{m=-\infty}^{+\infty} \sum\limits_{n=-\infty}^{+\infty} (d^m)^* C^{nm} e^n \text{ and}$$

$$<\varphi|T_C\psi> = \sum\limits_{m=-\infty}^{+\infty} \sum\limits_{n=-\infty}^{+\infty} (d^m)^* C^{nm} e^n = <T_C\varphi|\psi>. \tag{42}$$

b) Because of $T_C: s_C \to S_C$, $H_C: s_C \to s_C$ and $H_C: S_C \to S_C$ one has at first $[T_C, H_C]: s_C \to S_C$ so that the expectation values $<\varphi|[T_C, H_C]|\varphi>$, with $\varphi \in s_C$ and $\varphi(u) = U(u)\varphi \in s_C$ are well defined.

$<\varphi|[T_C, H_C]|\varphi> =$

$\tau^{-1} \int\limits_{-\tau/2}^{+\tau/2} du <\varphi|U^+(u)P_C U(u)H_C|\varphi> - \tau^{-1} \int\limits_{-\tau/2}^{+\tau/2} du <\varphi|H_C U^+(u)P_C U(u)|\varphi> =$

$\tau^{-1} \int\limits_{-\tau/2}^{+\tau/2} du <\varphi|U^+(u)P_C H_C U(u)|\varphi> - \tau^{-1} \int\limits_{-\tau/2}^{+\tau/2} du <\varphi|U^+(u)H_C P_C U(u)|\varphi> =$

$\tau^{-1} \int\limits_{-\tau/2}^{+\tau/2} du <\varphi(u)|P_C H_C|\varphi(u)> - \tau^{-1} \int\limits_{-\tau/2}^{+\tau/2} du <\varphi(u)|H_C P_C|\varphi(u)> =$

$i\tau^{-1} \int\limits_{-\tau/2}^{+\tau/2} du \, d/du <\varphi(u)|P_C|\varphi(u)> =$

$i\tau^{-1}<\varphi(+\tau/2)|P_C|\varphi(+\tau/2)> - i\tau^{-1}<\varphi(-\tau/2)|P_C|\varphi(-\tau/2)>. \tag{43}$

By (22) this yields

$$i\tau^{-1}[\tau + <\varphi(-\tau/2)|P_C|\varphi(-\tau/2)>] - i\tau^{-1}<\varphi(-\tau/2)|P_C|\varphi(-\tau/2)> = i. \qquad (44)$$

As a consequence, the symmetric

$K := i[T_C, H_C] + 1$ satisfies $<\varphi|K|\varphi> = 0$ for all $\varphi \in s_C$. (45)

Assume now that $\varphi, \psi \in s_C$ do exist so that $<\varphi|K|\psi> \neq 0$. Then $<\varphi|K|\psi>$ can be written as $<\varphi|K|\psi> = e^{i\alpha}|<\varphi|K|\psi>|$ with $\alpha \in [0, 2\pi[$.

The element $\varphi + e^{i\beta}\psi \in s_C$, where $\beta$ is any parameter, satisfies now

$$0 = <\varphi + e^{i\beta}\psi|K|\varphi + e^{i\beta}\psi> = e^{-i\beta}<\psi|K|\varphi> + e^{+i\beta}<\varphi|K|\psi> = \qquad (46)$$

$e^{+i(\alpha+\beta)}|<\varphi|K|\psi>| + e^{-i(\alpha+\beta)}|<\varphi|K|\psi>| = 2\cos(\alpha+\beta)|<\varphi|K|\psi>|$

so that $\cos(\alpha+\beta) = 0$.

The choice $\beta := -\alpha$ for example leads to a contradiction. Therefore

$<\varphi|K|\psi> = 0$ for all $\varphi, \psi \in s_C$. (47)

Let now be $K\varphi = \sum_{n=-\infty}^{+\infty} d^n \varphi_C(\tau^n) \in S_C$ with $\varphi \in s_C$. Because $\varphi_C(\tau^m) \in s_C$, yields

$0 = <\varphi_C(\tau^m)|K|\varphi> = d^m$ for all m so that $K\varphi = 0$ for all $\varphi \in s_C$. Summarizing

$[T_C, H_C] = i$ on $s_C$.

$U(t) = \exp(-itH_C)$ yields

$U^+(t)T_C U(t) = T_C + it[H_C, T_C] = T_C + t$, so that

$T_C U(t) = U(t)T_C + tU(t)$. This is the assertion

c) $d/dt <\varphi(t)|T_C|\varphi(t)> = -i<\varphi(t)|[T_C, H]|\varphi(t)> = 1$ so that

$<\varphi(t)|T_C|\varphi(t)> = t + <\varphi(0)|T_C|\varphi(0)>$. (48)

Because of (48) it remains to show that $<\varphi_C(0)|T_C|\varphi_C(0)> = 0$.

From (28) and (29) one has

$$<\varphi_C(0)|T_C|\varphi_C(0)> = C^{00} = \tau^{-1} \sum_{n=-\infty}^{+\infty} \int_{-\tau/2}^{+\tau/2} du \, |c^{0n}(u)|^2 \, \tau^n. \qquad (49)$$

By (06), (07) one gets for all $n \in \mathbb{Z}$

$$\int_{-\tau/2}^{+\tau/2} du \, |c^{0n}(u)|^2 = \int_{-\tau/2}^{+\tau/2} du \, |c^{0n}(-u)|^2 = \int_{-\tau/2}^{+\tau/2} du \, |c^{n0}(u)|^2 = \int_{-\tau/2}^{+\tau/2} du \, |c^{0\,-n}(u)|^2, \qquad (50)$$

so that $C^{00} = 0$ because of $\tau^{-n} = -\tau^n$.

d) The time invariance of σ(H_C) is trivial. The time invariance of σ(T_C) follows immediately from [T_C, U(t)] = tU(t).

In analogy to a well-known method one gets with (k > 0, φ ∈ s_C)

$$\Omega(k) := T_C - \langle\varphi|T_C|\varphi\rangle + ikH_C - ik\langle\varphi|H_C|\varphi\rangle \tag{51}$$

the term $\langle\Omega(k)\varphi|\Omega(k)\varphi\rangle = \text{Var}(T_C) + k^2 \text{Var}(H_C) - k \geq 0$, where $\Omega(k)\varphi \in S_C$. From this follows $k^{-1} \text{Var}(T_C) + k \text{Var}(H_C) \geq 1$. The left-hand side is minimal if $k = \sigma(T_C)/\sigma(H_C)$ and yields then $\sigma(T_C)\sigma(H_C) \geq \frac{1}{2}$. ∎

The concept of an IQC forbids the existence of eigenstates of $H_C$ and of $T_C$ in $s_C$ because $\langle\varphi|[T_C, H_C]|\varphi\rangle = 0$, if $\varphi \in s_C$ is such an eigenstate, in contradiction to (38). This non-existence of eigenstates is not only a formal result but has a physical background:

Eigenstates of $H_C$ cannot "age" so that they don't carry any time information: Let namely $\varphi \in s_C$, $\|\varphi\| = 1$, be any eigenstate of $H_C$ with the eigenvalue h so that $H_C\varphi = h\varphi$ and $\varphi(t) := U(t)\varphi = \exp(-iht)\varphi$. Then for all t

$$\langle\varphi(t)|T_C|\varphi(t)\rangle = \langle\varphi(0)|T_C|\varphi(0)\rangle. \tag{52}$$

Eigenstates of $T_C$ don't allow the time evolution at all.

Proof: Let $T_C\varphi = t_\varphi\varphi$ with $t_\varphi \in \mathbb{R}$, $\varphi \in s_C$ and $\|\varphi\| = 1$. One gets first $U(t)T_C\varphi = t_\varphi U(t)\varphi$, furtheron $(T_C - t)U(t)\varphi = t_\varphi U(t)\varphi$ and $T_C\varphi(t) = (t + t_\varphi)\varphi(t)$ where $\varphi(t) := U(t)\varphi$. In this way, $\varphi(t)$ is an eigenstate of $T_C$ to any t. Because of the symmetric $T_C$, eigenstates with different eigenvalues are orthogonal. Therefore $\|\varphi(t+dt) - \varphi(t)\|^2 = \langle\varphi(t+dt) - \varphi(t)|\varphi(t+dt) - \varphi(t)\rangle = +2$ for any $dt \neq 0$ so that the derivation $\lim[\varphi(t+dt) - \varphi(t)](dt)^{-1}$ for $dt \to 0$ does not exist. As a consequence, $\varphi(t)$ cannot satisfy Schrödinger's equation $H\varphi(t) = i\, d/dt\, \varphi(t)$, so that the time evolution of φ is impossible.

By the way, not only $s_C$ but also $S_C$ contains no eigenstates of $H_C$. Proof:

Assume that $\varphi \in S_C$, $\|\varphi\| = 1$, is an eigenstate of $H_C$ with the eigenvalue h so that $H_C\varphi = h\varphi$, and $U(t)\varphi = \exp(-iht)\varphi$. As an element of $S_C$, $\varphi$ may be written as $\varphi = \sum_{n=-\infty}^{+\infty} d^n \varphi_C(\tau^n)$ where

$$|d^m| = |\langle\varphi_C(\tau^m)|\varphi\rangle| = |\langle\varphi_C(\tau^m)|U^+(\tau^n-\tau^m)U(\tau^n-\tau^m)\varphi\rangle| = \qquad (53)$$
$$|\langle\varphi_C(\tau^n)|\exp[-ih(\tau^n-\tau^m)]\varphi\rangle| = |\langle\varphi_C(\tau^n)|\varphi\rangle| = |d^n|$$

for all m, n $\in \mathbb{Z}$, so that $d^n = 0$ for all n as a consequence of $|d^n| \to 0$ for $|n| \to \infty$. This contradicts the assumed existence of an eigenstate.

## 5. The time measurement by an IQC.

Let now be given an IQC C and a closed physical system D with a Hamiltonian $H_D$ so that the time evolution of D is determined by a Schrödinger curve $\psi(t)$, $t \in \mathbb{R}$, in a state space $S_D$. It is to discuss, how possible mechanisms of the interaction between C and D work and to what extent C measures the time difference of any prescribed initial and final states of D.

To describe this interaction, C and D are combined to a whole system C+D with the Hamiltonian $H_{CD} := H_C + H_D$ and a state space $S_{CD}$ which contains $s_C$ and $S_D$ as subspaces. The time evolution of C+D is then given by $\varphi_{CD}(t) := \varphi_C(t) + \psi(t) \in S_{CD}$ defined by $H_{CD}$. In general, C is disturbed by D (and vice versa). This disturbance is caused by nonvanishing projections of $\psi(t)$ into $s_C$ as follows:

Let C be given by $\varphi_C(t) \in s_C$ and the projection of $\psi(t)$ into $s_C$ by $\psi_C(t) \in s_C$. Then $\chi_C(t) := \varphi_C(t) + \psi_C(t) \in s_C$ is no Schrödinger curve of $s_C$ in general because the restriction of $H_D$ does not work as a Hamiltonian in $s_C$. The space $s_C$ namely admits $H_C$ as the only Hamiltonian because of $U(t)s_C = s_C$ with $U(t) = \exp(-itH_C)$. Therefore a condition of compatibility is introduced:

C and D are **compatible**, if the restriction of the Hamiltonian $H_D$ to the space $s_C$ is $H_C$.

Then $\chi_C(t)$ is Schrödinger curve in $s_C$, and (04) is satisfied in spite of the interaction between C and D. But the elements $\varphi_C(\tau^n)$ which defined C are now to be replaced by the $\chi_C(\tau^n)$, $n \in \mathbb{Z}$, which are not pairwise orthogonal in general (according to (03)). Thus C is no longer an ideal quantum clock. One has now the following cases:

### 5.1 No interaction between C and D:

Then $s_C \cap S_D = \{0\}$, and a suitable state space of C+D is the direct sum $s_C \oplus S_D$. Both systems run independently (and are compatible), and the Schrödinger curve is $\varphi_C(t) \oplus \psi(t)$, with $\|\varphi_C(t)\| = 1$ and $\psi(t) \in S_D$. The Hamiltonians $H_C$, $H_D$ operate separately on $s_C$ resp. $S_D$. The set of all these states is $\{\varphi_C(t) \oplus \psi(t) | t \in \mathbb{R}\}$, where the pairs $\varphi_C(t) \oplus \psi(t)$ are "coupled" by the same value of t. Let now be given any two states $\varphi_C^1 \oplus \psi^1$ and $\varphi_C^2 \oplus \psi^2$ from this set. $\psi^1$ may be an initial state of the system D, $\psi^2$ a final state. Then, the coupled $\varphi_C^1$, $\varphi_C^2$ yield the expectation values $<\varphi_C^i|T_C|\varphi_C^i>$, i = 1, 2, of the time operator of C. These $<\varphi_C^i|T_C|\varphi_C^i>$ are to be ordered to the corresponding $\psi^i$. The difference $|<\varphi_C^2|T_C|\varphi_C^2> - <\varphi_C^1|T_C|\varphi_C^1>|$ is independent of the chosen zero-point of the time given by $\varphi_C(0)$ and is taken as the duration of the process in question delivered by C. Because of the coupling $\varphi_C(t) \oplus \psi(t)$, the relation $<\varphi_C(t)|T_C|\varphi_C(t)> = t$, equ. (41), yields then t as the expectation value of the time operator also with respect to $\psi(t)$. Summarizing, the time measurement of D plays completely in $s_C$ and all features of an isolated C, especially the time-energy uncertainty relation, can be applied to C+D also.

**5.2 Weak interaction between C and D:**

It is again assumed that a state space $S_{CD}$ exists which contains $s_C$ and $S_D$ as subspaces. Provided the condition of compatibility, $\chi_C(t) := \varphi_C(t) + \psi_C(t) \in s_C$ is now the modified Schrödinger curve of C. The $\chi_C(\tau^n)$, $n \in \mathbb{Z}$, replace the $\varphi_C(\tau^n)$ of C, whereas the time operator $T_C$ (26) and the operator $P_C$ (21) are unchanged. Unlike the $\varphi_C(\tau^n)$, the changed elements $\chi_C(\tau^n)$ are not pairwise orthogonal in general. In this way, C gets now a non-ideal, but still usable clock, if the interaction between C and D is not too strong.

Introducing the set of pairs $\{(\chi_C(t), \psi(t)) | t \in \mathbb{R}\}$, coupled by the same value of t again, one may choose certain states $(\chi_C^i, \psi^i)$, i = 1, 2, from this set. $\psi^1$ may be an initial state, $\psi^2$ a final state of D. The resp. expectation values $<\chi_C^i|T_C|\chi_C^i>$ are now defined as the time values of the corresponding $\psi^i$ of D. As above, one gets then $|<\chi_C^2|T_C|\chi_C^2> - <\chi_C^1|T_C|\chi_C^1>|$ as the duration of the process of D, measured by C.

The just discussed mechanisms show that in extreme cases C may be disturbed by D in such a way that C gets completely unusable and, vice versa, D gets by C changed beyond recognition.

## Appendix: A note to the "theorem of Pauli".

The argument of Pauli against any (selfadjoint) time operator T exploits the properties of the formal unitary operator $\exp(+ikT)$ with any real parameter k, especially the equation $\exp(-ikT)H\exp(+ikT) = H + k$, which follows from the assumed identity $[T, H] = i$.

Accordingly, in the IQC-theory the operator

$$V(k) := \exp(+ikT_C): s_C \to S_C \tag{54}$$

to any fixed real k is to be discussed. First of all, as a consequence of the domain $s_C$ of $T_C$, the terms $T_C^2$, $T_C^3$ and so on, after all V(k), are not defined on $s_C$ in general, so that the maximal domain $D_{V(k)} \subseteq s_C$ of V(k) may be the trivial subspace $\{0\}$ of $s_C$ for certain k. At least, the common domain of all V(k) is the trivial subspace of $s_C$:

$$D_V = \{0\}, \text{ where } D_V := \bigcap D_{V(k)} \text{ for all k.} \tag{55}$$

Proof:

Let be $\varphi \in D_V$, $\|\varphi\| = 1$ and $\varphi_k := V(k)\varphi$. Then $\varphi_k \in S_C$ for all k. In Lemma 1 c, it is proven that a maximal energy W exists so that all expectation values $\langle\varphi|H_C|\varphi\rangle$ with $\varphi \in S_C$ and $\|\varphi\| = 1$ satisfy

$$|\langle\varphi|H_C|\varphi\rangle| \leq W. \tag{56}$$

As elements of $S_C$, $\varphi$ as well as $\varphi_k$ satisfy
$|\langle\varphi|H_C|\varphi\rangle| \leq W$ and $|\langle\varphi_k|H_C|\varphi_k\rangle| \leq W$. On the other hand
$\langle\varphi_k|H_C|\varphi_k\rangle = \langle\varphi|V(-k)H_C V(+k)|\varphi\rangle = \langle\varphi|H_C|\varphi\rangle + k$
because of $\exp(-ikT_C)H_C\exp(+ikT_C) = H_C + [-ikT_C, H_C]$, so that

$$|\langle\varphi|H_C|\varphi\rangle + k| = |\langle\varphi_k|H_C|\varphi_k\rangle| \leq W \text{ for all k.} \tag{57}$$

Contradiction. Therefore, $\varphi \in D_V$ with $\|\varphi\| = 1$ cannot exist, so that $\varphi = 0$. ∎